# qGAUSSIAN MODEL OF DEFAULT

Yuri A. Katz
S&P Capital IQ, 55 Water Str., New York, NY 10040, USA

***Abstract***. *We present the qGaussian generalization of the Merton framework, which takes into account slow fluctuations of the volatility of the firm's market value of financial assets. The minimal version of the model depends on the Tsallis entropic parameter q and the generalized "distance to default". The empirical foundation and implications of the model are illustrated by the study of 645 North American industrial firms during the financial crisis, 2006 - 2012. All defaulters in the sample have exceptionally large q > 3/2, corresponding to unusually fat-tailed unconditional distributions of log-asset-returns. Using Receiver Operating Characteristic curves, we demonstrate the high forecasting power of the model in prediction of 1-year defaults. Our study suggests that the level of complexity of the realized time series, quantified by q, should be taken into account to improve valuations of default risk.*

**Introduction.**

Structural models of default typically assume that the volatility of the issuer's financial assets can be described by the deterministic function of time (constant in the simplest case). This approach fails to recognize large intermittent bursts of the volatility of the stock price and, hence, the company's market value of assets, $V$. Evidently, the future value of the volatility of financial assets cannot be known today. It may be much lower or higher than the spot value of the volatility at the valuation time. As a result, the firm can be mistakenly classified as less or more risky than it truly is.

In the seminal Merton model (1974) the evolution of the logarithm of $V$ is modeled by the Brownian motion with constant coefficients of drift and diffusion, which leads to the log-normal distribution of $V$. The model further assumes that the debt of an issuer, $D$, is time invariant. The debt must be repaid at the maturity date $T$ and default may happen only if $V(T) \le D$. As a result, the cumulative probability of default (PD), as seen at the estimation time, $t = 0$, is defined by the CDF of the conditional Gaussian distribution of $\ln V(t=T)/D$ evaluated at the "default point", $V(T) = D$, and is equal to the *cumulative normal distribution* function of the "distance-to-default" (DTD). The Merton DTD is mainly determined by the inverse leverage ratio of the issuer at the valuation time, $\ln V(0)/D$, measured in the units of the standard deviation of the percentage daily changes of $\ln V$ ("log-asset-returns") in a time interval between the time of estimation and $T$. The value of the firm's DTD characterizes the proximity of the default point and is widely adopted as the key measure of default risk in financial industry (Duffie and Singleton, 2003; Lando, 2004). For recent econometric studies relating small DTDs (< 2) to a high risk of default see, eg, Löffler and Maurer (2008) and Flannery *et al.* (2012). It is well known, however, that due to "thin-tales" of the Gaussian distribution, substitution of the DTD larger than 3, which is common for investment-grade issuers, into the cumulative normal distribution function leads to severe underestimation of the forecasted 1-year PDs and short-term credit spreads. Generally, fluctuations of volatility violate the fundamental i.i.d. approximation (globally fluctuations of $V$ are not identically distributed) and, hence, the Gaussian underpinning of the Merton framework.

Although the traditional approximation of the time-invariant volatility is not valid, it may work as the starting point from which the more adequate description of the non-trivial stochastic behavior of financial time series can be derived. Following the insight of Praetz (1972), we decouple the Merton framework from its Gaussian foundation. The derived closed-form expression for the term structure of the cumulative PD takes into account the entire distribution of the inverse variance of the issuer's log-asset-returns at the valuation time, not merely the spot value. In the case of the Gamma distributed inverse variance, the likelyhood of default is determined by the CDF of the fat-tailed *qGaussian* distribution (Tsallis, 2009). The result depends on the *generalized DTD*, which preserves the simple intuition of the Merton DTD, however, is well defined even if the variance of the qGaussian distribution is divergent or undefined. The minimal version of the model contains only two parameters: the generalized Merton DTD and the Tsallis entropic parameter, $1 \leq q \leq 3$. For $q \to 1$, related to the normal statistics, the new approach replicates results of traditional structural models. For $q > 1$, corresponding to complex systems with non-additive entropy, the model leads to much higher 1-year PD for investment grade issuers than valuations based on the assumption of the constant volatility.

The empirical foundation and implications of the model are illustrated by the study of the time series of daily log-asset-returns of 645 North American industrial firms, including 44 defaulters, during the financial crisis, 2006 - 2012. We use both the direct proxy and iterative implied methods to estimate the firm's market value of assets. First, we present the detailed results for two investment-grade issuers: United Parcel Service (UPS) and United Technology (UTX). For both companies the unconditional qGaussian distribution fits well the empirical distribution of log-asset-returns, whereas the stochastic volatility of log-asset-returns has very long relaxation times (more than 50 trading days). For both issuers the qGaussian generalization of the traditional Black and Cox (1976) model of default leads to a significant increase of the forecasted 1-year PD. Next, we extend the analysis to the rest of the sample and report that all defaulters have exceptionally large entropic parameters $q > 3/2$, corresponding to unusually fat-tailed distributions. We use the method of Receiver Operating Characteristic (ROC) curves and demonstrate the high forecasting power of the qGaussian model in predicting 1-year defaults in the sample. Our study suggests that the degree of complexity of the realized time series characterized by the value of the Tsallis entropic parameter should be taken into account to improve valuations of the issuer's credit risk.

**The qGaussian generalization of the Merton framework.**

In the classic Merton model (1974), the default point is assumed to be constant, $D = const$ and the time evolution of the conditional PDF, $p_\beta(x, t | x_0, t_0 = 0) := p_\beta(x, t | x_0)$, to find the realization $x = \ln V / D$ at the moment $t$, given $x_0 = \ln V_0 / D$ at the beginning of the time interval $(0, t]$, is approximated by the unbounded diffusion process with the constant variance per unit of time, $\sigma^2 = 1/\beta$, and the constant effective drift, $m = \mu - \sigma^2 / 2$. The solution of this problem for some fixed value of the variance and the deterministic initial condition $p_\beta(x, t = 0 | x_0) = \delta(x - x_0)$ is described by the following Green function:

$$p_\beta(x, t | x_0) = \sqrt{\frac{\beta}{2\pi t}} \exp\left[-\beta \frac{(x - x_0 - mt)^2}{2t}\right]. \tag{1}$$

Since, by assumption, default can happen only if $V(T) \leq D$, the cumulative survival probability of the obligor $\Omega_\beta^{(M)}(T|x_0)$, as seen at the valuation time, $t_0 = 0$, is determined by the integral over $x$ of the "snapshot" of Eq.(1) at the maturity time $t = T$:

$$\Omega_\beta^{(M)}(T|x_0) = \int_0^\infty p_\beta(x, T | x_0) dx = \frac{1}{2}\left[1 + erf\left(\frac{x_0 + mT}{\sqrt{2T}} \sqrt{\beta}\right)\right], \tag{2}$$

which immediately yields the celebrated Merton result for the cumulative PD:

$$PD_\beta^{(M)}(T|x_0) = 1 - \Omega_\beta^{(M)}(T|x_0) = \Phi\left[-dd_\beta(T)\right]. \tag{3}$$

Here $\Phi[z] = [1 + erf(z/\sqrt{2})]/2$ is the cumulative normal distribution function, $erf(z)$ is the error function,

$$dd_\beta(T) = [\ln(1/R_0) + m T]\sqrt{\beta/T} \tag{4}$$

is the Merton DTD for some fixed value of the precision of the Gaussian distribution, $\beta$, and $R_0 = \ln D/V_0$ is the leverage ratio of an issuer at $t_0 = 0$.

Similarly, in the first passage type of structural models pioneered by Black and Cox (1976), the temporal behavior of $p_\beta(x, t | x_0)$ is approximated by the Brownian motion with a constant coefficient of diffusion. The key novel element is the account of early defaults that may happen at any interim point in time, $0 < t \leq T$. Specifically, in the Black-Cox model, default occurs at the first hitting time, whenever the diffusive path of $\ln V$ hits the pre-specified absorbing default boundary, which is equal to $\ln D$ or some lower threshold (the debt level that triggers default). The Green function of this problem is well known, see eg, Chandrasekhar (1943):

$$p_\beta(x, t | x_0) = \sqrt{\frac{\beta}{2\pi t}} \exp\left[\beta m(x - x_0) - \beta m^2 t/2\right]\left\{\exp\left[-\beta \frac{(x - x_0)^2}{2t}\right] - \exp\left[-\beta \frac{(x + x_0)^2}{2t}\right]\right\}$$

which leads to the classic Black-Cox result

$$PD_\beta^{(BC)}(t|x_0) = \Phi\left[-\frac{x_0 + m t}{\sqrt{t}} \sqrt{\beta}\right] + \exp(-2m x_0 \beta) \Phi\left[-\frac{x_0 - m t}{\sqrt{t}} \sqrt{\beta}\right]. \tag{5}$$

As expected, the account of early defaults leads to generally higher PD at $t = T$ forecasted by Eq.(5) than by the Merton formula, Eq.(3). In particular, if we disregard the effective drift, $m = 0$, than the cumulative PD estimated by the Black-Cox Eq.(5) is two times higher than the one predicted by Eq.(3):

$$PD_\beta^{(M)}(T|x_0) = \Phi\left[-d\tilde{d}_\beta(T)\right], \tag{6a}$$

$$PD_\beta^{(BC)}(t = T|x_0) = 2\Phi\left[-d\tilde{d}_\beta(T)\right]. \tag{6b}$$

Note that in these expressions the term structure of the cumulative PD is determined by the "simplified" Merton DTD (with $m = 0$):

$$d\tilde{d}_\beta(t) = x_0\sqrt{\beta/t} = \frac{\ln(1/R_0)}{\sigma\sqrt{t}}. \tag{7}$$

This measure of credit risk is very popular in financial industry, for it helps to reduce sampling errors related to estimations of the drift at relatively short time scales.

Now let us introduce an uncertainty related to the fluctuating volatility of the issuer's log-asset-returns. In the next section we demonstrate that empirically the relaxation time of the volatility is much longer than the relaxation time of log-asset-returns. This behavior is consistent with "volatility clustering" observed on equity markets, see Mantegna and Stanley (2000) and Bouchaud and Potters (2000). In this regime, the variance can be considered as *locally* constant over the time period, which is sufficiently long for the fast variable to locally obey the i.i.d. approximation. This assumption significantly simplifies the description of the temporal evolution of the fast variable. In this limit, the local solution of the unbounded diffusion of *x* is Gaussian, conditional upon the stochastic realization of its precision $\beta$, which is determined by the distribution, $f(\beta)$. Consequently, in the continuous space, the *global* (compound) Green function of the unbounded diffusion with randomly distributed diffusion coefficient, $p(x,t|x_0)$, is obtained by marginalization over $\beta$ of the infinite set of locally normal conditional distributions $p_\beta(x,t|x_0)$:

$$p(x,t|x_0) = \int_0^\infty p_\beta(x,t|x_0) f(\beta) d\beta . \qquad (8)$$

This approach to modeling of financial time series was pioneered by Praetz (1972) in early seventies. The "mixture of distributions" hypothesis proposed by Praetz (1972) is very close to the methodology employed in the modern field of "superstatistics", which has recently been a topic of increasing interest in physics of complex systems and finance (Beck and Cohen, 2003; Gerig *et al.*, 2009; Van der Straeten and Beck, 2009; Vamos and Craciun, 2010; Takaishi, 2010; Camargo et al., 2013; Katz and Tian, 2013; Katz and Tian, 2014).

It follows from Eqs.(2), (3), and (8) that the effect of the slowly fluctuating volatility of log-asset-returns on the term-structure of the cumulative PD at the estimation time $t_0 = 0$ can be taken into account by simply averaging of the conditional $PD_\beta(t|x_0)$ over the distribution $f(\beta)$:

$$\overline{PD}(t|x_0) = \int_0^\infty PD_\beta(t|x_0) f(\beta) d\beta = 1 - \int_0^\infty dx \int_0^\infty p_\beta(x,t|x_0) f(\beta) d\beta \quad . \qquad (9)$$

According to Eq.(9), the likelihood of bankruptcy at the valuation time depends upon the default triggering event and the model of stochastic motion of *x* represented by the local Green function, $p_\beta(x,t|x_0)$, as well as the realized quasi-static distribution $f(\beta)$. Empirical analysis of the time-series of the estimated market value of financial assets (Katz and Tian, 2014) suggests that the reciprocal of the variance of daily log-asset-returns is Gamma distributed:

$$f(\beta) = \frac{b^a}{\Gamma(a)} \beta^{a-1} \exp(-b\beta) \quad . \qquad (10)$$

Here *a* and *b* represent the shape and the rate (inverse scale) parameters of the Gamma distribution. Note that the equilibrium Gamma distribution of $\beta$ may result from different "microscopic" feedback mechanisms driving volatility fluctuations (Hull and White, 1987; Biro and Rosenfeld, 2007; Golan and Gerig, 2013). Henceforth, we assume that the realized distribution $f(\beta)$ is described by the empirically relevant and theoretically appealing Gamma distribution. Substitution of Eqs.(6) and (10) into Eq.(9) after integration over $\beta$ yields the following generalization of the traditional results:

$$\overline{PD}^{(M)}(t=T|x_0) = \frac{1}{2}I_{\frac{1}{1+x_0^2/(2bT)}}(a,\frac{1}{2}) = \frac{1}{2}I_{\frac{1}{1+(q-1)d\tilde{d}^2(T)/2}}(\frac{3-q}{2q-2},\frac{1}{2}) \quad , \quad (11a)$$

$$\overline{PD}^{(BC)}(t|x_0) = I_{\frac{1}{1+x_0^2/(2bt)}}(a,\frac{1}{2}) = I_{\frac{1}{1+(q-1)d\tilde{d}^2(t)/2}}(\frac{3-q}{2q-2},\frac{1}{2}) \quad . \quad (11b)$$

Here $I_z(m,n)$ is the regularized incomplete Beta function (Zelen and Severo, 1972), which replaces the cummulative normal distribution [cf., Eqs.(6a) and (6b)] and we use the following parameterization: $q=(2a+3)/(2a+1)$ and $\tilde{\beta}=(2a+1)/(2b)$ on the far right hand side of Eqs.(11a) and (11b) to relate these results to the Tsallis entropic parameter $q$ and the scale parameter $\tilde{\beta}$ of the leptokurtic qGaussian distribution of log-asset-returns. It is important to mention here that under certain constrains the unconditional qGaussian distribution maximizes the Tsallis form of non-additive entropy (Tsallis, 2009). It provides the natural generalization of the Gaussian distribution for $q>1$ and appears in a wide spectrum of complex systems, including financial markets (Borland, 2002; Ausloos and Ivanova, 2003; Tsallis *et al.,* 2003; Borland and Bouchaud, 2004; Friedman *et al.,* 2007; Tsallis, 2009). For relatively small entropic parameters, $q<5/3$, the variance per unit of time of the qGaussian distribution has the finite value $\tilde{\sigma}_q^2 = 2/[(5-3q)\tilde{\beta}]$. It is divergent for $5/3 \leq q < 2$ and is undefined for $2 \leq q < 3$. The tails of the qGaussian distribution decrease as power laws with the exponent $\sim 2/(q-1)$.

It is straightforward to verify that these results can be obtained by changing the order of integration in Eq.(9): first, take the average of $p_\beta(x,t|x_0)$ with respect to the realized probability distribution $f(\beta)$, then integrate over $x$ from negative infinity to zero. The first step replaces the conditional Gaussian distribution with the conditional qGaussian distribution, while the second step yields the CDF of this distribution, see Appendix for details. Note that, up to normalization of the scale parameter, Eq.(11b) is the exact transformation of the formula obtained by Katz and Tian (2013). The results presented here contain $I_z(m,n)$ with the argument $z \leq 1$. This representation of the regularized incomplete Beta function has been implemented in popular statistical tools, eg, R and is convenient for applications. As expected, if the Tsallis entropic parameter $q \to 1$, the Gaussian distribution is restored and the 2-parametric Eqs.(11a) and (11b) reduce to the conventional 1-parametric results, Eqs.(6a) and (6b). This regime corresponds to the large shape parameter $a \gg 1$. In this limit, the distribution $f(\beta)$ is sharply peaked about the mean, $\beta_0 = a/b = \int_0^\infty \beta f(\beta) d\beta$, which corresponds to the case of a constant volatility.

Notably, equations (11a) and (11b) depend on the generalized DTD:

$$d\tilde{d}(t) = x_0\sqrt{\tilde{\beta}/t} \quad . \quad (12)$$

Similarly to the simple Merton DTD, Eq.(7), it is proportional to the inverse leverage ratio of the issuer at the time of estimation. On the other hand, it depends on the shape parameter of the qGaussian distribution of log-asset-returns. Hence, the generalized DTD is well defined even if the variance of the observed fat-tailed qGaussian distribution of log-asset-returns is divergent or undefined. For relatively small entropic parameters, far from the critical value of 5/3, we may compare values of the simple Merton DTD, Eq.(7), and the generalized DTD, Eq.(12):

$$\frac{d\tilde{d}}{d\tilde{d}_\beta} = \sqrt{\frac{\tilde{\beta}}{\beta}} = \frac{\sigma}{\tilde{\sigma}}\sqrt{\frac{2}{5-3q}} \quad . \tag{13}$$

Assuming that for $q < 5/3$, $\sigma \approx \tilde{\sigma}$, the generalized DTD should be slightly bigger than the simple Merton DTD. Nevertheless, due to fatter tails of the underlining conditional qGaussian distribution, the relevant CDF (the regularized incomplete Beta function) may yield significantly higher PD than the cumulative normal distribution function.

Figure 1 shows the 3D plot of the 1-year $\overline{PD}^{(BC)}$ forecasted by the qGaussian generalization of the Black-Cox model of default, Eq.(11b). The likelihood of default is calculated for different values of $q$ and the generalized DTD. It is easy to see from Fig.1 that for issuers with high credit quality, even a small increase in $q$ leads to a sharp growth of the 1-year PD, which reflects the growing probability of extreme movements in the time-series of log-asset-returns. Notably, the generalized DTD and the 1-year $\overline{PD}^{(BC)}$ are well-defined and finite in the area of $5/3 \leq q < 2$, corresponding to the unconditional qGaussian distribution with the divergent variance. Figure 1 clearly demonstrates that for $q > 1$ the estimated 1-year PD follows the relatively smooth, *power-law decay* with the growth of $d\tilde{d}$. It is easy to see from Fig.1 that the influence of large $q$'s is less pronounced in the area of small $d\tilde{d}$.

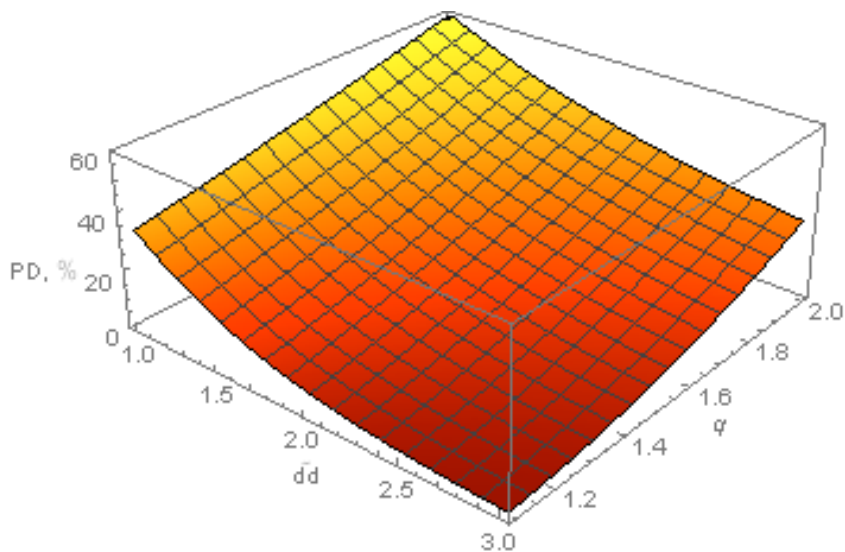

Figure 1. (Color online) 3D plot of the 1-year PD forecasted by the 2-parametric qGaussian generalization of the Black-Cox model of default.

These visual observations can be augmented by the analysis of the relevant asymptote of Eq.(11b). In the region of large generalized DTD, $(q-1)d\tilde{d}^2(t) >> 1$, we obtain the following asymptotic

$$\overline{PD}^{(BC)}(t|x_0) \approx \frac{2(q-1)^{\frac{q-2}{q-1}}}{(3-q)C_q}\left(\frac{2t}{\tilde{\beta}x_0^2}\right)^{n/2} = \frac{2(q-1)^{\frac{q-2}{q-1}}}{(3-q)C_q}\left[\frac{\sqrt{2}}{d\tilde{d}(t)}\right]^n. \quad (14)$$

Here $n = (3-q)/(q-1)$ is the power-low exponent and the $q$-dependent factor $C_q$ is determined as follows, see Appendix, Eq.(A3): $C_q = \frac{1}{\sqrt{q-1}}B\left[\frac{1}{2},\frac{3-q}{2(q-1)}\right]$, where $B[1/2,\bullet]$ is the Beta function. Note that the relevant asymptote of Eqs.(6b) goes to zero extremely fast: $PD_\beta^{(BC)}(t|x_0) \propto \exp[-d\tilde{d}_\beta^2(t)]$ and becomes negligibly small at relatively large 1-year Merton DTDs, $d\tilde{d}_\beta(t=1) > 3$. Empirically, large values of DTD are typically estimated for issuers with investment grade credit ratings. These firms have relatively small, but yet non-zero likelihood of bankruptcy within one year from the estimation time. We shall return to this important point in the next section.

In the opposite limit, $(q-1)d\tilde{d}^2(t) << 1$, the asymptote of Eq.(11b) follows the ordinary $\sim t^{-1/2}$ prescription:

$$\overline{PD}^{(BC)}(t|x_0) \approx 1 - \frac{1}{C_q}\sqrt{\frac{2\tilde{\beta}x_0^2}{t}} = 1 - \sqrt{2}\frac{d\tilde{d}(t)}{C_q}. \quad (15)$$

It is straightforward to see from the definition of the generalized DTD, Eq.(12), that this regime is inevitable at long time scales as the role of the fat tails of the conditional qGaussian distribution [see Appendix, Eqs.(A2) and (A3)] is diminished with the growth of time. Furthermore, according to Eq.(15), for issuers with large leverage ratio at the estimation time, $x_0 = \ln(1/R_0) << 1$, the close proximity of the default point diminishes the role of slow volatility fluctuations in estimations of the 1-year PD. Finally, it is easy to see that if $q \to 1$, the estimate Eq.(15) reduces to the relevant asymptote of the simplified Black-Cox formula, Eq.(6b). Note that the maximum value of the cumulative multi-period PD in both Merton, Eq.(6a), and the generalized Merton, Eq.(11a), models is equal to 50%, which makes both of them inapplicable for long-term forecasting and for financial distressed firms in the vicinity of the default point. On the other hand, due to the more realistic default triggering condition of the Black-Cox model, the realistic maximum value of the likelihood of bankruptcy, 100%, is achieved in both versions of the model. Therefore, hereafter, we focus only on the qGaussian generalization of the Black-Cox model, Eq.(11b).

**The empirical foundation and implications of the qGaussian model.**

To empirically verify the key assumptions and implications of the qGaussian generalization of the Merton framework, we collect the data set, which consists of 645 North American industrial public companies (GICS 20) through the financial crisis, from July 11, 2006 to June 21, 2012. The sample includes 44 issuers that have been defaulted through the period of observation. For the detailed illustration we pick two US industrial firms: United Parcel Service (UPS) and United Technology (UTX). Both companies keep the investment-grade credit ratings through the whole period of observations (1500 trading days). In particular, UPS was rated 'AAA', then downgraded to 'AA-' by S&P Ratings Services, whereas UTX has the 'A' rating, assigned by S&P Ratings Services, between July 11, 2006 and June 21, 2012.

Implementation of structural models requires the accounting balance sheet data and the company's stock market information. There are several estimation methods of the firm's market value of assets $V$ and the default point $D$. For instance, in forecasting 1-year PD the default point is often determined as the sum of the short-term debt and the half of the long-term debt of the issuer (Crosbie and Bohn, 2003). One of the drawbacks of this approach is related to ambiguity of the definition of short- and long-term liabilities. Here we use the book value of total liabilities of the issuer at the date of valuation as the proxy for $D$. This measure has no ambiguity in the definition and reflects an existence of covenants that may force the issuer to serve both short- and long-term debts when its financial situation deteriorates. Obviously, this specification may lead to an upward bias in the estimate of $D$ (not all debt maturing in one year). Therefore, as the alternative, in our estimations we also use 80% of the book value of total liabilities as the proxy for $D$. To estimate the market value of the issuer's assets we, first, follow the simple and rather popular in the corporate finance literature direct proxy method. The method assumes that $V$ is equal to the sum of the firm's end-of-trading-day market capitalization $E$ and $D$. The book value of company's total liabilities is reported quarterly and can be interpolated on a daily basis (Eom *et al.,* 2004; Bharath and Shumway, 2008; Flannery *et al.,* 2012). Second, we apply the more sophisticated method based on the iterative algorithm (Crosbie and Bohn, 2003; Vassalou and Xing, 2004). In this method, the sum of $E$ and $D$ represents only the initial guess value of $V$. The fast converging iterative procedure employs the Black-Scholes call-option formula to obtain the daily time series of *implied* asset values.

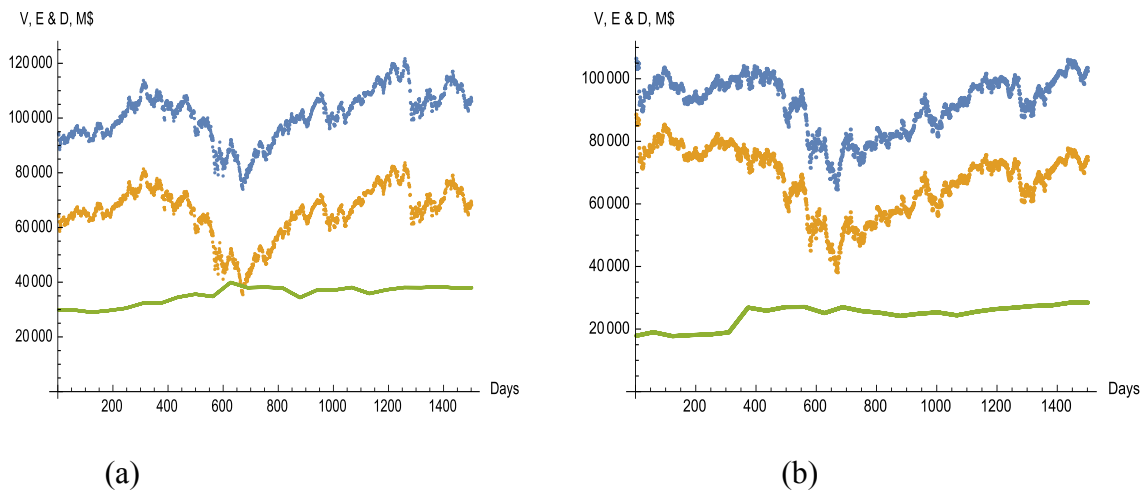

(a)  (b)

Figure 2. (Color on line) Time series (bottom – up) of the accounting book value of total liabilities, the market capitalization, and the market value of assets estimated by the direct proxy method: (a) UTX and (b) UPS (July 11, 2006 - June 21, 2012). *Source: S&P Capital IQ*

We extract the book value of the firm's total liabilities as well as its end-of-trading-day market capitalization from the S&P Capital IQ database and smooth quarterly jumps in the reported total liabilities of issuers by linear interpolation. With the daily time series of $V$ in place, we compute the daily log-asset-returns defined as the forward daily change in the logarithm of a firm's market value of assets $\upsilon_{i,j} = \ln(V_{i,j}/V_{i-1,j})$, where $i$ and $j$ denote the trading day and the company, respectively. Analysis of the time series of $\upsilon$ is the necessary first step that paves the way towards model-based valuations of the likelihood of default. If, for example, the unconditional distribution of $\upsilon$ is close to Gaussian, the standard diffusion process may be a good approximation for the stochastic evolution of the issuer's $\ln V$. In this case, one can use the traditional structural models to estimate the issuer's PD. Figures 2(a) and 2(b) show the time series of the accounting book value of the total debt, the market capitalization, and the market value of assets estimated by the direct proxy method, $V = E + D$, for UTX and UPS between July 11, 2006 and June 21, 2012. These plots exhibit relatively small changes in $D$ and rather large bursts in the time-evolution of $E$ and $V$ for during the period of observation. They clearly demonstrate that for both companies, fluctuations of $E$ determine the stochastic behavior of $V$. In fact, we verify that due to the relatively low debt level and its stability, the unconditional distributions of daily log-asset-returns $\upsilon_{i,j} = \ln(V_{i,j}/V_{i-1,j})$ and log-leverage-returns $r_{i,j} = \ln(R_{i,j}/R_{i-1,j})$ are practically indistinguishable. We use the MatLab implementation of the maximum likelihood estimate (MLE) method to fit both unconditional Gaussian and qGaussian distributions to the observed time series of log-asset-returns. The "Q-Q" plots on Fig.3 allow a visual comparison of the distributions of $\upsilon$ for UTX and UPS. On the *x*-axis we plot the quantiles of the fitted theoretical distributions of log-asset-returns, whereas the quantiles of the empirical distributions are on the y-axis. It is easy to see from Figs.3(a) and 3(b) that for UTX and UPS the observed distribution and the fitted qGaussian distribution of log-asset-returns are similar; the points of the Q–Q plots lie very close to the $y = x$ line. On the other hand, comparison of the quantiles of the observed distribution of $\upsilon$ with the quantiles of the fitted normal distribution on Figs.3(c) and 3(d) shows the strong dissimilarity of empirical and theoretical results in the tail areas. We augment these graphical tests with multiple statistical 'goodness-of-fit' tests. Notably, Anderson-Darling, Cramer-von Mises, Kolmogorov-Smirnov, and Person $\chi^2$ tests strongly reject the Gaussian as the null hypothesis and support the qGaussian distribution as the alternative hypothesis at 5% accuracy. Note that application of the iterative implied method to estimations of $V$ practically does not change these results.

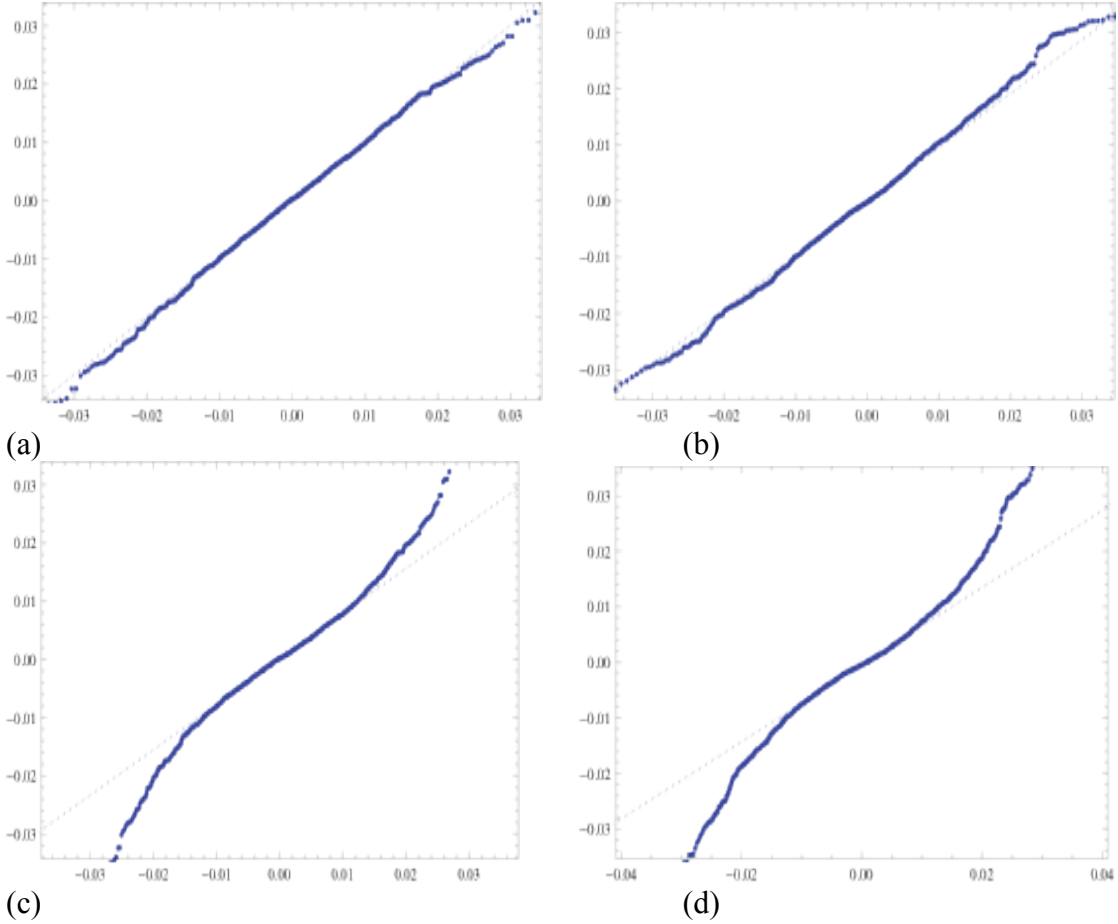

Figure 3. Q-Q plots of the distributions of daily log-asset-returns estimated by the direct proxy method. (a), (b) Quantiles of the fitted qGaussian distribution are on the x-axis for UTX and UPS, respectively. (c), (d) Quantiles of the fitted Normal distribution are on the x-axis for UTX and UPS, respectively.

Importantly, the behavior of the auto-correlation functions of daily log-asset-returns demonstrates an instantaneous (1 trading day) loss of memory, which is not shown here. On the other hand, the time series of $|\upsilon|$ and $\upsilon^2$ related to the volatility and variance of log-asset-returns, exhibit strong autocorrelation with more than 50 trading days memory lag, $h$. The relevant auto-correlation functions (ACF) of $|\upsilon|$ for UTX and UPS

$$ACF(|\upsilon|_j, h) = \sum_{i=1}^{N-h}(|\upsilon|_{i,j} - \hat{m}_j)(|\upsilon|_{i+h,j} - \hat{m}_j) / \sum_{i}^{N}(|\upsilon|_{i,j} - \hat{m}_j)^2 \qquad (16)$$

are shown on Figs.4(a) and 4(b). In Eq.(16), $\hat{m}_j$ is the mean of the data set consisting of $N=1500$ absolute values of daily log-asset-returns of the company $j$. It is easy to see from these plots that the stochastic dynamics of the volatility of log-asset-returns is rather slow, much slower than fluctuations of $\upsilon$. Notably, the time evolution of the ACFs of $|\upsilon|$ and $|r|$ are practically indistinguishable. This conclusion holds for both specifications of the default point used in our study. As expected, these findings are consistent with the statistical patterns observed in the time series of stock prices (Mantegna and Stanley, 2000; Bouchaud and Potters, 2000). This observation supports the key assumption of the qGaussian model: the random variance of the process can be considered as *locally* constant over the time period that is sufficiently long for the fast variable $v$ to locally obey the probabilistic law governed by the standard diffusion equation.

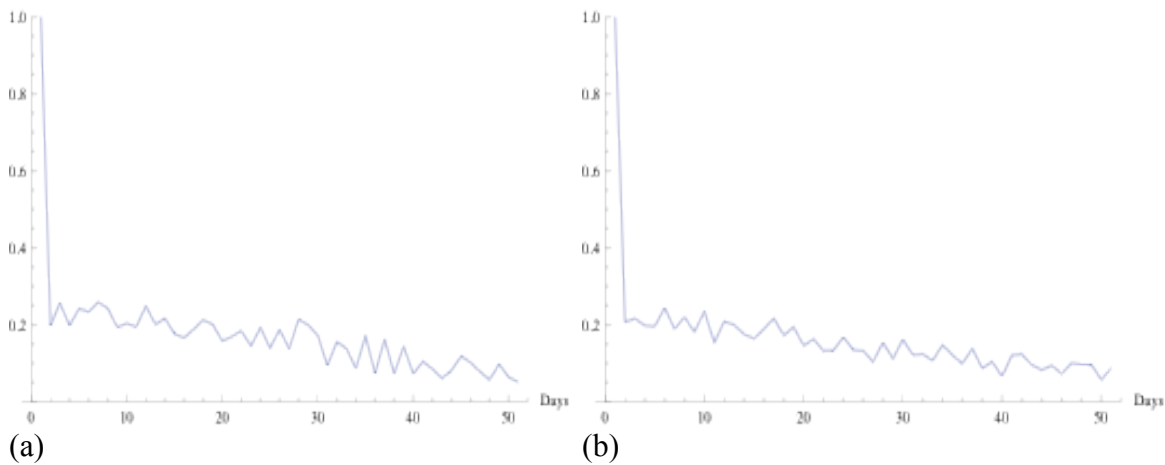

(a) (b)

Figure 4. Auto correlation functions of the absolute value of daily log-asset-returns estimated by the direct proxy method. (a) UTX and (b) UPS. See text for details.

Moving from the analysis of unconditional distributions towards the estimation of the 1-year PD, we use both the direct proxy and iterative implied methods to estimate the firm's market value of assets. For each company we use the rolling window of past 250 trading days and apply the MLE method to fit parameters $q_{i,j}$ and $\tilde{\beta}_{i,j}$ of the unconditional qGaussian distribution to the realized time-series of $v_{i,j}$ on the daily basis. Figures 5 and 6 show the daily evolution of the fitted entropic factor as well as the simple Merton and the generalized DTDs of UTX and UPS during the *peak* of the financial crisis, 2008 – 2009. We see that the parameters are mostly stable and their behavior is quite similar, but not identical for UTX and UPS. Both issuers have relatively small values of $q < 3/2$, with generalized DTDs always larger than the simple Merton DTD, which is consistent with our estimate, see Eq.(13). Plugging the fitted parameters $q_{i,j}$ and $\tilde{\beta}_{i,j}$ into Eqs.(6b), (11b), and (12) we derive the relevant 1-year PDs forecasted by the Black-Cox model and its qGaussian generalization. Figure 7 shows the outcome of these calculations. It is easy to see that for both investment-grade issuers characterized by relatively small entropic parameters and large DTDs account of the realized distribution of inverse variance at the valuation time leads to a significant increase of the forecasted 1-year PD. The change of the estimation of the firm's market value of assets from the direct proxy to the iterative implied method virtually does not change this plot. On the other hand, if other things are equal, the 20% decrease in the default barrier $D$ leads to ~ 50% decrease in the estimated value of 1-year PD in both models under consideration. This reduction leads to the more reasonable values of the forecasted 1-year PD by the qGaussian model of default during the peak of the financial crisis, which for an average global issuer initially rated 'AA' or 'A' should be ~ 0.4% (Vazza and Kraemer, 2013).

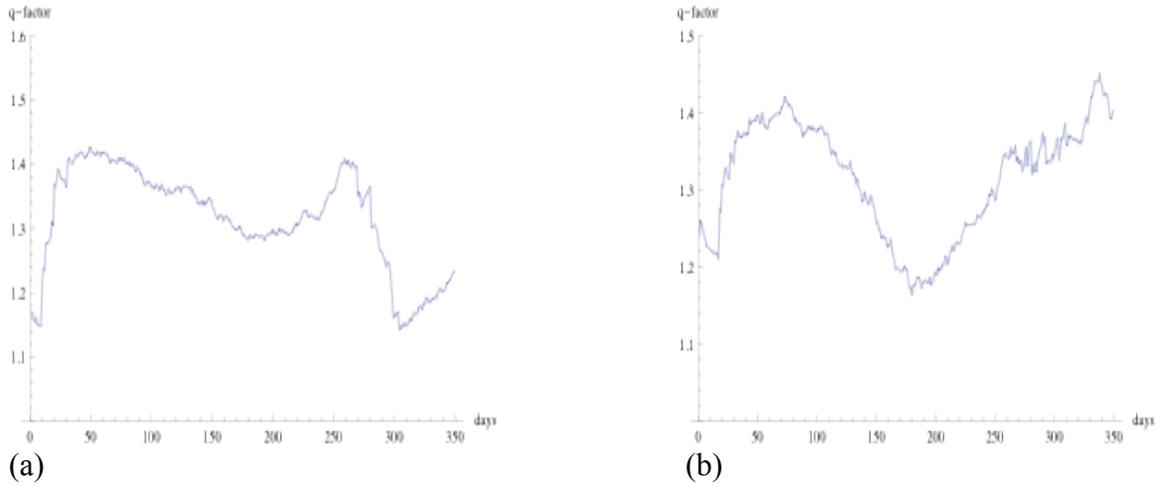

Figure 5. Evolution of the fitted entropic parameters of the qGaussian distributions of the log-asset-returns estimated by the direct proxy method. (a) UTX and (b) UPS. The beginning of the days-count corresponds to the past 550 trading days from the start of our observations on July 11, 2006 and is approximately related to the beginning of the financial crisis, 2008 – 2009. See text for details.

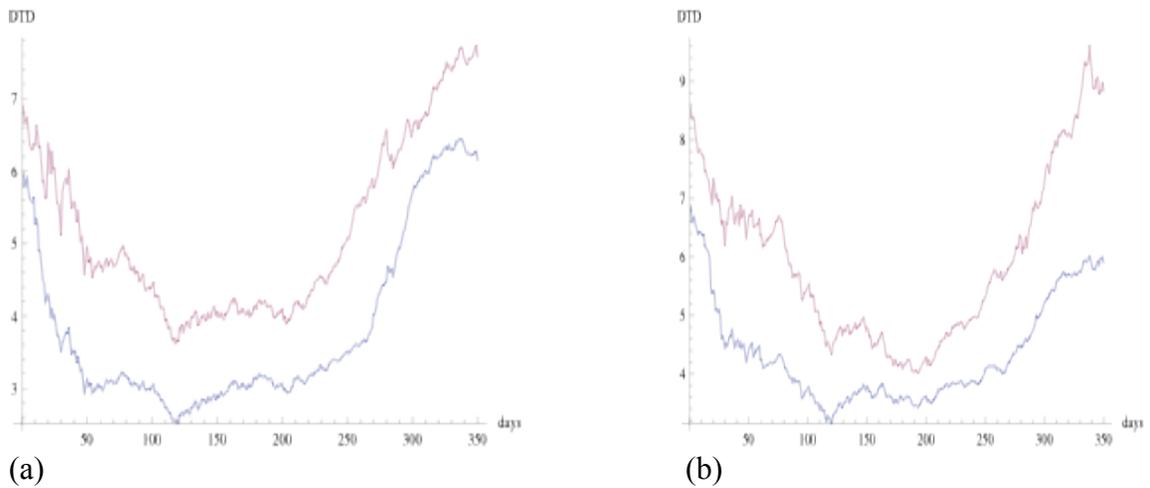

Figure 6. (Color online) Evolution of the estimated 1-year DTDs. The upper curves represent generalized DTD, whereas the lower curves correspond to simple Merton DTD (a) UTX and (b) UPS. The beginning of the days-count corresponds to the past 550 trading days from the start of our observations on July 11, 2006. See text for details.

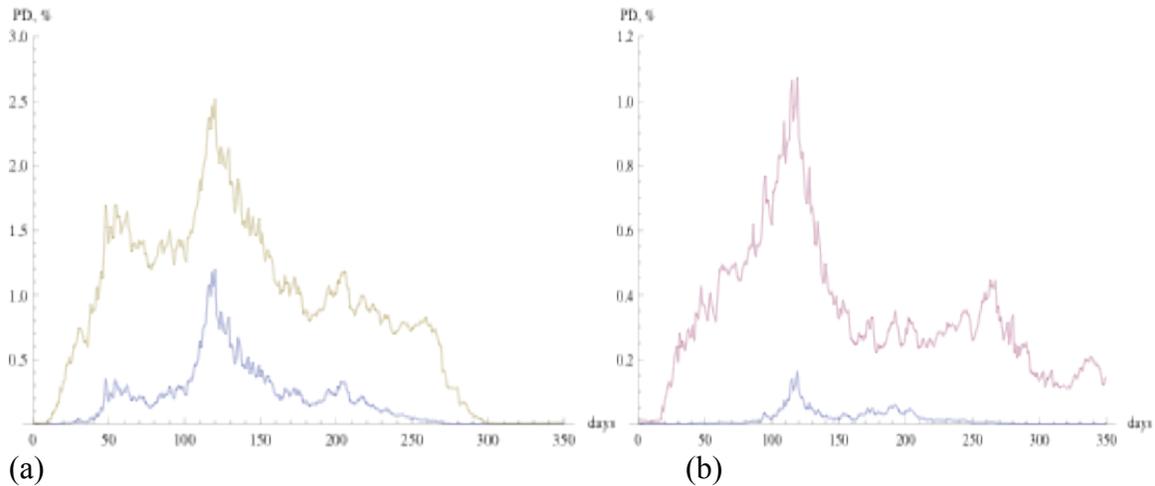

(a)  (b)

Figure 7. (Color online) Evolution of the 1-year PD estimated on the daily basis by the Black-Cox (lower curves) and qGaussian (upper curves) models of default. (a) UTX and (b) UPS. The beginning of the days-count corresponds to the past 550 trading days from the start of our observations on July 11, 2006 and is approximately related to the beginning of the financial crisis, 2008 – 2009. See text for details.

Extending the analysis to the rest of the sample, we find that the time series and statistical patterns of daily log-asset-returns and log-leverage-returns obtained with either of the very different estimation methods – direct proxy and iterative implied - are very similar. Moreover, this observation does not depend on the selection of the default point used in this paper. The close proximity of results suggests that for North American industrial corporations, with usually small proportion of liabilities in the capital structure, the time series of the market value of assets are dominated by fluctuations of the equity value and the Merton assumption, $D = const$, is the good approximation. Nevertheless, for the vast majority of 645 North American industrial firms between July 11, 2006 and June 21, 2012, multiple statistical goodness of fit tests reject the Gaussian distribution as the null hypothesis and support the qGaussian unconditional distribution of daily log-asset-returns as the alternative hypothesis with 5% confidence, see also Katz and Tian (2014). This observation does not depend on the method of estimation of firm's market value of assets and is consistent with findings on equity markets.

The histogram on Fig.8 shows the distribution of the fitted entropic parameters $q_j$ among issuers in the sample. It shows that most of non-defaulted firms have qGaussian distributions of log-asset-returns with finite variance, $1 < q < 5/3$. Interestingly, all defaulters have large values of entropic parameters, $q > 3/2$, with 84% of them having $q \geq 5/3$, which corresponds to divergent or undefined variance of the distribution. Qualitatively, large entropic parameters reflect the high probability of extreme daily fluctuations of log-asset-returns, revealed by the fat-tails of the qGaussian distribution, which evidently are related to an elevated risk of default. Notice, however, that not every issuer with a large entropic parameter files for bankruptcy protection during 6 years of our observations. Both the value of $q$, reflecting the complexity of the realized time-series of log-asset-returns and the issuer's log-leverage ratio at the valuation time determine the risk of default.

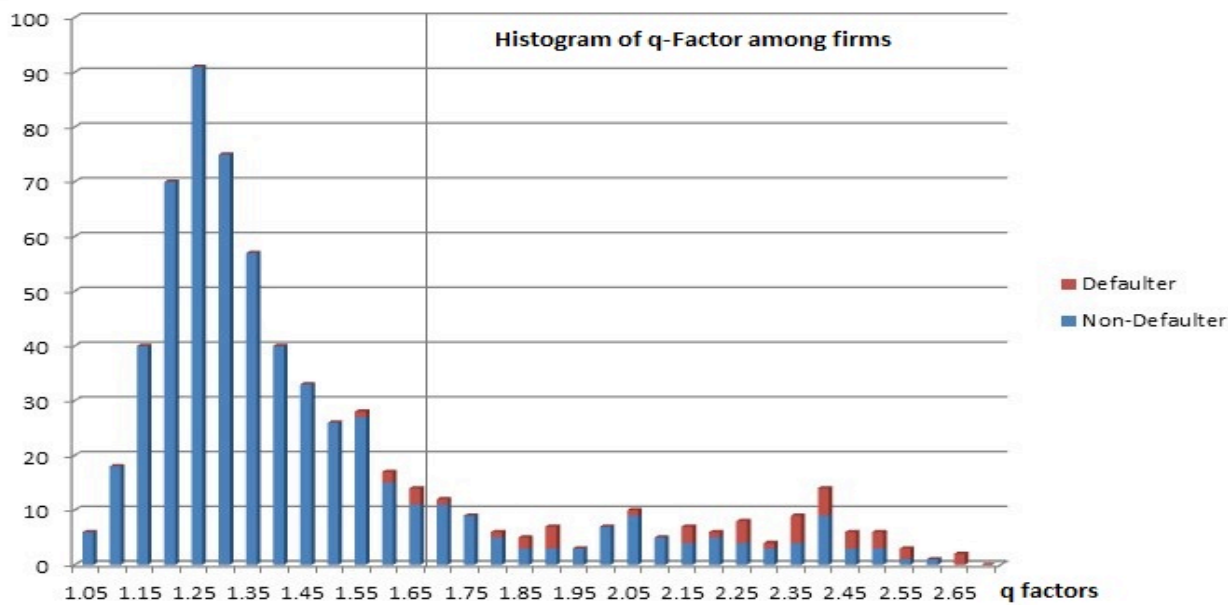

Figure 8. (Color online) Distribution of the fitted entropic parameters among issuers: 645 North American industrial companies between July 11, 2006 and June 21, 2012. The vertical line corresponds to $q = 5/3$. See text for details.

    To quantify the predictive power of the qGaussian generalization of the traditional Black-Cox model we use the method of ROC curves. This method is broadly used to illustrate the performance of any binary classifier system in machine learning and data mining (Trevor *et al.*, 2009) as well as in measuring the forecasting power of default risk models (Engelman *et al.*, 2003). To generate the ROC curve we calculate quantiles of the "true positive" predictions (*y*-axis) and "false positive" forecasts or Type I errors (*x*-axis). The values on the *x*-axis are ordered from 0 to 1 by the false positive rate, which is the number of defaulters classified by the PD model as low-risk (PD < threshold value) divided by the total number of defaulters. A perfect credit risk model is the one that classified all defaulted firms in the worst ranking category, with the ROC curve going sharply up, from 0 to 1, along the y-axis of true positives. Obviously, a model with no predictive power will generate the 45 degrees line and have the Area Under the Curve (AUC) of 1/2. The value of AUC is commonly used to quantify the predictive power of different models. Under this metric, the ideal model has AUC = 1.

    The ROC/AUC methodology is very sensitive to the quality of the data set. Therefore, we remove companies with market capitalization bellow $150M at the beginning of our observations on July 11, 2006 from our initial sample. We also drop defaulters that have less than one year of history. We ended up with the sample consisting of 361 non-defaulted issuers and 29 defaulters. To improve statistics, we randomly choose non-defaulters, such that the number of selected firms is equal to the number of bankruptcies in each year of our observations between 2007 and 2012. For each selected company *j* and for each available trading day *i* we use the rolling window of past 250 trading days and the MLE method to fit parameters $q_{i,j}$ and $\tilde{\beta}_{i,j}$ of the unconditional qGaussian distributions to the realized time-series of $\upsilon_{i,j}$. Then, for both defaulters and

randomly selected non-defaulters we compute the 1-year $d\tilde{d}_{i,j}(t=1)$ and $\overline{PD}_{i,j}^{(BC)}(t=1)$ according to the qGaussian generalization of the Black-Cox model, Eq.(11b). We set the default point $D_j$ of the firm $j$ to the book value of total liabilities and calculate the ROC curves for both estimation methods of the market value of assets $V_{i,j}$ described above. We repeat this exercise 100 times, by randomly choosing non-defaulters per model's specification, to get a distribution of AUC values and determine the model performance based on the average AUC value across simulations. Figure 9 shows typical sets of the derived ROC curves calculated with and $V_{i,j}$ obtained by the iterative implied method. We find that the forecasting power of the qGaussian generalization of the Black-Cox model in predicting 1-year PD is very high. The average AUC values for both estimation methods of $V_{i,j}$ are very close, between 0.96 and 0.97. The high forecasting power of the model is consistent with the theoretical analysis presented in the previous section and empirical studies see, eg, Flannery *et al.* (2012), which demonstrate that for companies *approaching default*, the high leverage ratio and the high realized spot volatility at the estimation time are the key predictive factors of bankruptcy. For issuers within a year from default, account of the historical fluctuations of the volatility of log-asset-returns does not significantly improve the forecasting power of the model. Note, however, that the traditional approach to valuation of the 1-year PD is conceptually inapplicable for distressed companies, which typically have very fat-tailed distributions of log-asset-returns with divergent or undefined second moment.

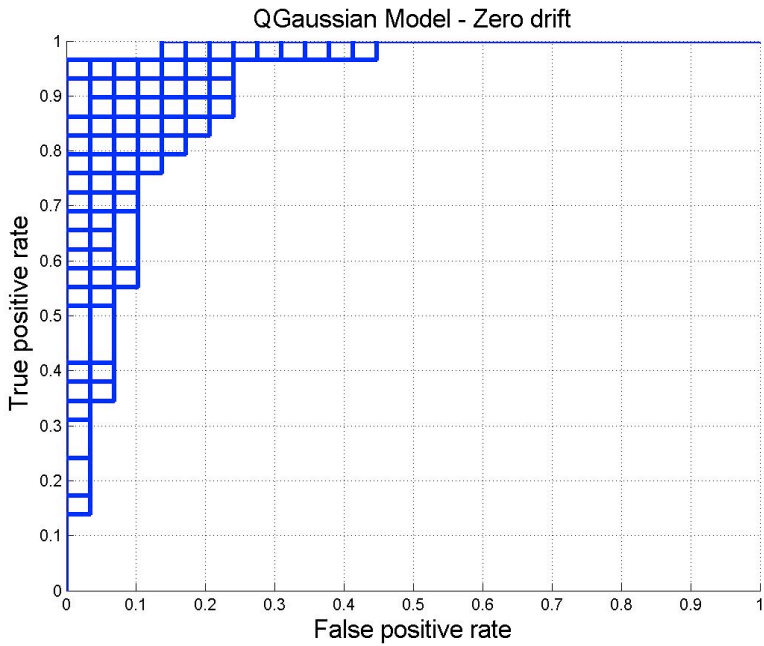

Figure 9. ROC curves calculated for 1-year PDs estimated with market values of assets obtained by the iterative implied method for the sample consisting of 361 non-defaulted firms and 29 defaulters, between 2007 and 2012. See text for details.

## 3. Summary.

We analyze the influence of uncertainty related to slowly fluctuations of the volatility of the firm's market value of assets, estimated from the stock market and accounting data, on the likelihood of default. Our approach closely follows the insight of Praetz (1972), who applied the concept of the compound probability distribution to explain leptokurtic empirical distributions of stock price returns. In our formulation of the problem, the global conditional qGaussian distribution of log-asset-returns replaces the conditional Gaussian distribution of the classic Merton framework. The resulting cumulative PD is determined by the proximity of the default point as well as the generally quasi-static realized distribution of the inverse variance of the issuer's log-asset-returns, $f(\beta)$, at the valuation time. The derived compact formulae, Eqs.(11a) and (11b), depend on the generalized Merton DTD, Eq.(12), and reveal the dependency of the forecasted cumulative PD on the complexity of the realized time series. The latter is characterized by either parameters $a$ and $b$ of the Gamma distribution of the precision of the Gaussian distribution, $\beta$, or equivalently, by the parameters $q$ and $\tilde{\beta}$ of the qGaussian distribution of log-asset-returns. For sharp, pointed distributions of $\beta$, corresponding to large shape parameters, $a \gg 1$, or small entropic factors, $q \cong 1$, the obtained expressions reproduce the traditional Merton and Black-Cox results. On the other hand, slow fluctuations of $\beta$ lead to an increased complexity of time series reflected by big $q$'s and fat-tails of the global unconditional qGaussian distribution of log-asset-returns. In this regime, the model forecasts much higher 1-year PD for investment grade issuers, which are relatively far from the default point at the moment of valuation.

We observe that most non-defaulters among 645 North American industrial companies between 2006 and 2012 have unconditional qGaussian distributions of log-asset-returns with finite variance, $1 < q < 5/3$. In contrast, all defaulted companies in the sample have unusually large values of the Tsallis entropic parameter, $q > 3/2$. We find that the Merton assumption of the time-invariant default point, $D = const$, is the good approximation for North American industrial firms, with usually small proportion of liabilities in the capital structure. We confirm that the volatility and variance of daily log-asset-returns have the long relaxation time (more than 50 trading days). This observation supports the main assumption of the qGaussian model: the random variance of the process can be considered as locally constant over the time period, which is sufficiently long for the fast variable (log-asset-returns) to *locally* obey the i.i.d. approximation.

In general, it is qualitatively clear that the predictive power of the model positively depends on the amount of information available at the valuation time. Therefore, knowledge of the realized distribution $f(\beta)$ at the moment of estimation of the likelihood of default is more valuable than its spot value. The simple generalization of the Merton framework presented here has a number of practical advantages. The qGaussian model is easy to implement. The model's parameters are stable, intuitive, and straightforward to calibrate. Notably, the generalized DTD has the simple intuition of the Merton DTD and is well defined even for empirically relevant qGaussian distributions of daily log-asset-returns with divergent or undefined variance, $q \geq 5/3$. The qGaussian extension of the Black-Cox model shows the high forecasting power in predicting 1-year defaults for issuers in our sample. The average AUC values for 1-year default forecasting with various estimation methods of the firm's market value of assets are very close and rather high, between 0.96 and 0.97. Note that firms approaching default within one year typically have very fat-tailed distributions of log-asset-returns. For these issuers the traditional approach to valuation of the 1-year PD is conceptually inapplicable.

The main conclusion of our study: the complexity of the realized time series of the issuer's log-asset-returns should be taken into account *in addition* to the proximity of the default point. It would be of interest to apply the qGaussian model under the risk-neutral probability measure to arbitrage-free pricing of credit instruments. Note that under the risk-neutral probability measure the derived formulae are also applicable to valuations of "down-and-out" barrier options that cease to exist when the price of the underlying security hits a specific barrier price level. Naturally, the minimal 2-parametric version of the qGaussian model can be extended to achieve more realistic reflection of the stochastic dynamics of the firm's market value of assets, differentiate between the market-wide and idiosyncratic reasons of volatility fluctuations.

**APPENDIX**

Integration over $\beta$ from 0 to infinity of the product of the Green function of the unbounded diffusion, Eq.(1), with the Gamma distribution, Eq.(10), yields the global conditional PDF of the unbounded diffusion process with the slowly fluctuating coefficient of diffusion:

$$p(x,t \mid x_0) = \frac{1}{B(1/2, a)\sqrt{2bt}} \left[1 + \frac{(x-x_0)^2}{2bt}\right]^{-(a+1/2)}$$

$$= \frac{1}{B(1/2, \nu/2)\sqrt{\nu s^2 t}} \left[1 + \frac{(x-x_0)^2}{\nu s^2 t}\right]^{-(\nu/2+1/2)} \quad (A1)$$

Here $B(1/2, a) = \Gamma(1/2)\Gamma(a)/\Gamma(1/2+a)$ is the Beta function, $\Gamma(a)$ is the Gamma function, and use the following parameterization: $\nu = 2a$ and $s = \sqrt{b/a} = 1/\sqrt{\beta_0}$ on the far right hand side of Eq.(A1) to relate these results to the well known scaled Student *t*-distribution.

It is easy to see that, with yet another change of parameterization, either $q = (2a+3)/(2a+1)$ and $\tilde{\beta} = (2a+1)/(2b)$ or $q = (\nu+3)/(\nu+1)$ and $\tilde{\beta} = (\nu+1)/(\nu s^2)$, Eq.(A1) is identical to the qGaussian distribution, for values of the Tsallis entropic parameter within the interval $1 < q < 3$ (de Souza and Tsallis 1997):

$$p(x,t \mid x_0) = \frac{1}{C_q}\sqrt{\frac{\tilde{\beta}}{2t}} e_q[-\tilde{\beta}\frac{(x-x_0)^2}{2t}] \quad . \quad (A2)$$

Here $e_q(\bullet)$ is the qGaussian function and $C_q$ is the normalization *q*-dependent factor determined as follows

$$e_q\left[-\frac{\tilde{\beta}x^2}{2t}\right] = \left[1 + \tilde{\beta}(q-1)\frac{x^2}{2t}\right]^{-\frac{1}{q-1}}, \quad C_q = \frac{1}{\sqrt{q-1}} B\left[\frac{1}{2}, \frac{3-q}{2(q-1)}\right] \quad . \quad (A3)$$

It is straightforward to see that CDF of Eq.(A1) evaluated at the default point yields Eq.(11a).

**AKNOWLEDGEMENT**
The views expressed in this paper are those of the author, and do not necessary represent the views of S&P Capital IQ. The author is grateful to Viktor Gluzberg and Nikolai Shokhirev for helpful discussions and critical comments on a range of problems dealt with in this paper.**REFERENCES**.
Ausloos M. and Ivanova K. (2003). Dynamical model and nonextensive statistical mechanics of a market index on large time windows. *Phys. Rev. E,* **68,** 046122.
Beck C. and Cohen E. G. D. (2003). Superstatistics. *Physica* A, **322**, 267-275.
Biro T. S. and Rosenfeld R. (2007). Microscopic origin of non-Gaussian distributions of financial returns. *Physica* A, **387,** 1603-1612.
Bharath, S. T. and Shumway T. (2008). Forecasting default with the Merton Distance to Default model, *Review of Financial Studies,* **21**, 1339-1369.
Black F. and Cox J. C. (1976). Valuing corporate securities: Some effects of bond indenture provisions. *J. Finance*, **31**(2), 351-367.
Borland L. (2002). A theory of non-Gaussian option pricing. *Quantitative Finance*, **2**, 415-431.
Borland L. and Bouchaud J. P. (2004). A non-Gaussian option pricing model with skew. *Quantitative Finance*, **4**, 499-514.
Bouchaud J. P. and Potters M. (2000). *Theory of Financial Risks: From Statistical Physics to Risk Management.* Cambridge: Cambridge University Press.
Camargo S., Duarte Queiros S. M., and Anteneodo C. (2013). Bridging stylized facts in finance and data non-stationaries. *EPJ* B, **86,** 159-172.
Chandrasekhar S. (1943). Stochastic problems in physics and astronomy. *Rev. Mod. Phys.*, **15**, 1.
Crosbie P. and Bohn J. (2003). Modeling default risk. *Moody's KMV technical document*.
de Souza A. M. C. and Tsallis C. (1997). Student's t- and r-distributions: Unified derivation from an entropic variational principle. *Physica* A **236,** 52.
Duffie D. and Singleton K. J. (2003). *Credit Risk: Pricing, Measurement, and Management.* Princeton: Princeton University Press.
Eom, Y., Helwege J., and Huang J. (2004). Structural models of corporate bond pricing: an empirical analysis. *Review of Financial Studies,* **17**, 499-544.
Flannery M. J., Nikolova S., and Oztekin O. (2012). Leverage expectations and bond credit spreads. *J. Financial and Quantitative Analysis* (JFQA), **47**, 689-714.
Friedman C., Huang J., and Sandow S. (2007). A utility based approach to some information measures. *Entropy*, **9** (1) 1 - 26.
Gerig A., Vicente J., and Fuentes M. A. (2009). Model for non-Gaussian intraday stock returns. *Phys. Rev. E*, **80**, 065102(R).
Golan R. and Gerig A. (2013). A stochastic feedback model for volatility. http://arxiv.org/abs/1306.4975. Working paper.
Hull J. and White A. (1987). The pricing of options on assets and stochastic volatility. *J. Finance,* **42** (2), 281-300.
Katz Y. A. and Tian L. (2013). Q-Gaussian distributions of leverage returns, first stopping times, and default risk valuations. *Physica* A, **392,** 4989-4996.
Katz Yu. A. and Tian L. (2014). Superstatistical fluctuations in time series of leverage returns. *Physica* A, **405,** 326-331.18